\documentstyle[12pt]{article}
\input epsf
\newcommand{\semi}{;\hfil\break}

\newcommand{\e}[1]{\label{eq:#1}}
\newcommand{\ee}[1]{(\ref{eq:#1})}

\newcommand{\eq}{\begin{equation}}
\newcommand{\eqe}{\end{equation}}

\newcommand{\eqa}{\begin{eqnarray}}
\newcommand{\eqae}{\end{eqnarray}}

\newcommand{\del}{\partial}
\newcommand{\zb}{{\bar{z}}}

\begin{document}

\thispagestyle{empty}
\rightline{NSF-ITP-94-73}
\rightline{hep-th/9407031}

\vskip 3.5 cm

\begin{center}
{\large\bf COMBINATORICS OF BOUNDARIES IN STRING THEORY} \break

\vskip 2.5 cm
{\bf Joseph Polchinski}\footnote{Electronic address:
joep@sbitp.itp.ucsb.edu}

\vskip 0.7 cm
\sl
Institute for Theoretical Physics  \break
University of California  \break
Santa Barbara, CA 93106-4030  \break

\end{center}
\vskip 1.6 cm
\rm

\begin{quote}
{\bf ABSTRACT:}
We investigate the possibility that stringy nonperturbative effects appear
as holes in the world-sheet.  We focus on the case of Dirichlet string theory,
which we argue should be formulated differently than in previous
work, and we find that the effects of boundaries are naturally weighted by
$e^{-O(1/g_{\rm st})}$.
\end{quote}
\vskip1cm

\normalsize

\newpage

\section{Introduction}

It is essential to develop a nonperturbative understanding of string
theory.  The discovery of unusually large nonperturbative effects, of
order
$e^{-O(1/g_{\rm st})}$, is likely to be an important clue.  These effects
are found in matrix models and are expected more generally
from the large-order behavior of string perturbation
theory\cite{Shenker}, but they have no field-theory analog and their
nature is not in general known.
In this paper we suggest that the leading stringy nonperturbative
effects might make their appearance through holes (boundaries) in the
string world-sheet, and study in detail the case of Dirichlet boundary
conditions.  We show in particular that holes naturally have weight
$e^{-O(1/g_{\rm st})}$.

Dirichlet boundary conditions in string theory have been a subject of
frequent study.  Boundaries with Dirichlet conditions on all coordinates
have been considered as a source of partonic behavior in string
theory\cite{parton1,parton2,parton3} and as external probes on the
theory\cite{CMNP,CMNP2}.
Boundaries with Dirichlet conditions on some coordinates and Neumann on
others have been suggested to represent a form of
compactification\cite{Siegel} or an extended object in
spacetime\cite{DLP}.
The interpretation here is new---that they represent an example of the
sought-after $e^{-O(1/g_{\rm st})}$ stringy non-perturbative effects.
Our claim hinges on a treatment of the combinatorics of the boundaries
which differs from that previously assumed in the fully Dirichlet case,
though it is closely related to the work~\cite{DLP} on extended
objects.  We will argue that our combinatorics is that which arises
naturally from duality, and that it is necessary for the
Fischler-Susskind treatment of string divergences.  However, while we
will at times contrast our approach with others, we are not necessarily
asserting that the latter are incorrect.  These represent distinct
theories, with different physics, and the consistency of each (including
our own) is still an open issue.

In the next section we review the interpretation of mixed
Dirichlet/Neumann boundary conditions in terms of
extended objects, with emphasis on the combinatorics.
We then consider the fully Dirichlet case and
show how the $e^{-O(1/g_{\rm st})}$ weight arises.  We
discuss divergences and conformal anomalies, which are always important
constraints in string theory, and show that the Fischler-Susskind
mechanism operates.  We conclude with a brief discussion of the
implications, and suggest that the Dirichlet boundary is only one example
of a much larger class.

\section{D-Branes}

We begin by reviewing the critical bosonic string with Dirichlet
conditions on one coordinate $X^{25}$ and Neumann conditions on
the remaining $X^{\mu} \equiv
X^0, \ldots, X^{24}$.  This can be obtained by duality from the
more familiar fully Neumann theory by compactifying the Neumann
coordinate $\tilde X^{25}$ and taking the radius to
zero\cite{DLP,Gdual}.  To see this, consider some world-sheet path $C$
which begins at a boundary point $p_1$ and ends at any other boundary
point $p_2$. The Neumann coordinate $\tilde X^{25}$ and the dual
coordinate
$X^{25}$ are related by
\eq
\tilde X^{25}(z,\zb) = X_L^{25}(z) - X_R^{25}(\zb), \qquad
X^{25}(z,\zb) = X_L^{25}(z) + X_R^{25}(\zb).
\eqe
Then
\eqa
X^{25}(p_2) - X^{25}(p_1) &=& \int_C dz \,( \del_z X^{25} +
\del_\zb X^{25} ) \nonumber\\
&=& \int_C dz \,( \del_z \tilde X^{25} -
\del_\zb \tilde X^{25} ) \nonumber\\
&=& 0.
\eqae
The last equality holds because the line-integral is the total
$\tilde p^{25}$ momentum crossing $C$, and this is zero for all states
surviving in the $R \to 0$ limit.  We conclude that all string endpoints
are at the same value of $X^{25}$.
That is, if the world-sheet has several disconnected boundary
components, the field $X^{25}$ satisfies a Dirichlet boundary condition
with {\it the same} value on each component.  Let us for now define the
zero mode of $X^{25}$ such that the value is zero.

What is the physical interpretation?  Open strings endpoints lie
on the hyperplane $X^{25} = 0$, while in the rest of the dual
spacetime only closed strings are found.  The physics is that of a
closed string theory in interaction with a 24-dimensional extended
object, the `D(irichlet)-brane.'
In a relativistically invariant theory one does not expect
to find a rigid object, and indeed, the
shape of the D-brane is dynamical.  The vertex operator
\eq
\oint_B ds\, A(X^\mu) \del_n X^{25}, \e{vo}
\eqe
($n$ = normal, $t$ = tangential)
corresponds to fluctuations of the shape of the D-brane.  This is evident
from consideration of the boundary state: the normal derivative
$\del_n X^{25}$ is the canonical momentum for $X^{25}$, so the vertex
operator introduces an $X^\mu$-dependent shift of $X^{25}$.
The leading action for these fluctuations is the world-volume swept out by
the D-brane, as developed
in ref.~\cite{DLP}.
The vertex operator~\ee{vo} is dual to the gauge field vertex operator
$\oint_B ds\, A(X^\mu) \del_t \tilde X^{25}$
of the Neumann theory.

{}From cluster decomposition, one would expect that there should also
exist configurations with two or more D-branes.
Indeed, these arise from duality when Chan-Paton factors are included.
Introduce a Chan-Paton degree of freedom with states $a = 1, \ldots, N$.
The diagonal vertex operators
\eq
\lambda_{aa} \oint_B ds\, A_a(X^\mu) \del_n X^{25}, \e{vo2}
\eqe
(no sum on $a$) produce an $a$-dependent shift of $X^{25}$, so the
boundary shape depends on the state $a$ of the boundary: each of the
$N$ states corresponds to a {\it different} D-brane.\footnote
{The off-diagonal vertex operators correspond to open strings with
endpoints on different D-branes, so these are all massive when the
D-branes are separated.}  The path integral thus includes
\eq
\sum_{N=0}^\infty \left\{ \biggl(
\prod_{a=1}^N \int [dX^{25}_a] \biggr) \sum_{n=0}^\infty
\sum_{a_1,\ldots,a_n = 1}^N \right\}. \e{comb0}
\eqe
That is, for each $N$ sum over the number $n$ of world-sheet
boundaries and sum each of the $n$ Chan-Paton degrees of freedom
from 1 to $N$; then, integrate over the configurations of the
$N$ D-branes, and sum over the number $N$ of D-branes.  Summing
over $n$ and the Chan-Paton factors is equivalent to summing,
for each value of the Chan-Paton factor (each D-brane), over the number
of world-sheet boundaries lying in the given D-brane.
Thus, the sum~\ee{comb0} is equivalent to
\eq
\sum_{N=0}^\infty \prod_{a=1}^N \left\{ \int [dX^{25}_a]
\sum_{n_a=0}^\infty \right\}. \e{comb}
\eqe

Note that cluster decomposition in the dual theory requires a
sum over the number $N$ of Chan-Paton degrees of freedom.  This suggests
that string theories with different Chan-Paton groups should be regarded
as different states of a single theory, not a surprising result given the
increasing evidence for the unity of string theory.  Incidently, we have
been limited thus far to D-branes with the topology {\bf R}$^{24}$.  One
would expect other topologies, such as $S^{24}$, but these do not seem
to have any simple dual description.\footnote{Even D-branes of
topology {\bf R}$^{24}$ are problematic in the dual theory if they
become folded over so that $X^{25}(X^\mu)$ is multi-valued.}

\section{D-Instantons}

For the fully Dirichlet case, the boundary is at a single spacetime point
and so corresponds to an event, a D-instanton.
The sum~\ee{comb} becomes
\eq
\sum_{N=0}^\infty
\prod_{a=1}^N \left\{ \int [d^{26}X_a] \sum_{n_a=0}^\infty \right\}.
\e{comb2}
\eqe
That is, the functional integral over the configuration reduces to a
26-dimensional integral over the position of each D-instanton.  The
sum~\ee{comb2} differs from that considered in the partonic Dirichlet
theory\cite{parton1,parton2,parton3}
where the spacetime position of each world-sheet
boundary is integrated independently---equivalently, each $n_a$ is fixed
at $1$ for $a = 1, \ldots, N$.  Roughly speaking, with the
combinatorics~\ee{comb2}, the D-instanton is an extrinsic event in
spacetime, to which any number of world-sheet boundaries may attach,
while with $n_a \equiv 1$ each D-instanton is associated with exactly one
world-sheet boundary.

Consider the one-D-instanton amplitude $A_1$.  The sum~\ee{comb2} includes
world-sheet components which have Dirichlet boundaries at $X$ but are
otherwise disconnected; the leading such contribution would be the
disk amplitude with no vertex operators, denoted $<1>_{D_2}$.  The
amplitude with $\nu$ such disks includes a symmetry factor $1/\nu!$,
so the sum exponentiates,
\eq
A_1 = \exp( <1>_{D_2} + \ldots \,)\, A^{\rm connected}_1,
\eqe
where the ellipsis represents higher-order disconnected topologies.
A string world-sheet of Euler number $\chi$, with $m$ vertex
operators, is weighted by $g_{\rm st}^{m - \chi}$.  In particular, the
disk with no vertex operators has weight $g_{\rm st}^{-1}$, giving the
advertised $e^{-O(1/g_{\rm st})}$ (the sign in the exponent will
be obtained below).

For $m$ vertex operators, the leading contribution to the connected
amplitude, order $g_{\rm st}^{0}$, comes from $m$ disks each with a
single vertex operator,
\eq
A_1 = \exp( <1>_{D_2} + \ldots\,)
\int d^{26}X  \prod_{i=1}^m <V_i>_{D_2,X} + \ldots \ ,
\e{amp}
\eqe
with the position $X$ of the boundary
noted by a subscript where relevant.
Notice that momentum is not conserved on the individual world-sheet
components; it flows through the boundary and is only conserved in the
total process.  This amplitude can be summarized in terms of an
effective action
\eq
S_{\rm eff} \sim \int d^{26}X \exp\Bigl( <1>_{D_2} +
\sum_\alpha A_\alpha <V_\alpha>_{D_2,X} \Bigr) \e{seff}
\eqe
The sum $\alpha$ runs over all vertex operator types (including an
implicit momentum integration), and $A_\alpha$ is the corresponding
creation or annihilation operator,

The actual coefficient of $1/g_{\rm st}$ is of some interest.  The
results~\cite{tad} are readily extended to the Dirichlet case,
giving
\eqa
<1>_{D_2} &=& -\left(\frac{3}{16\pi}\right)^{3/2} \frac{2}{\pi^2}
\cdot (2\pi)^{13/2} \cdot \left(\frac{16\pi}{3}\right)^{3/2}
\frac{2^{-11/2}\pi^2}{\kappa_{26}} \cdot (2\pi \alpha')^6
\nonumber\\
&=& -\frac{2^8 \pi^{25/2} \alpha'^6}{\kappa_{26}}.
\eqae
In the first equality the four groups of terms, separated by dots,
are from the conformal Killing volume, the relation of
determinants on the disk to those on the sphere, the relation of the
sphere amplitude to the gravitational coupling, and a dimensional factor
restoring units.  This is for the oriented case; the unoriented would
have an additional $2^{-1/2}$.  Convert to an effective four-dimensional
coupling,
\eq
\kappa_{26}\ =\ V^{1/2} \kappa_4\ =\ v^{1/2} (4\pi^2 \alpha')^{11/2}
\kappa_4,
\eqe
where $V$ is the compactification volume and $v$ is the same in
units of the minimum volume of toroidal compactification.   Also,
$\kappa_4 = g_4 \alpha'^{1/2} /2$, taking the example of a level one
$SU(n)$ group.  In all,
\eq
<1>_{D_2} = -\frac{\pi^{3/2}}{4 v^{1/2} g_4}.
\eqe
In terms of the four-dimensional gauge coupling, the coeffecient is of
order~1 for moderate values of $v$.  This is comparable to the value
estimated by Banks and Dine by very different means\cite{BD}, and so is
consistent with their proposal that the stringy nonperturbative effects
can be large even while field-theoretic nonperturbative effects remain
quite small.

What is the effect of the D-instanton amplitude~\ee{amp}?  The main
previous interest in Dirichlet boundaries has been their partonic
scattering behavior.  Indeed, there is no Regge suppression in the
one-point amplitude on the disk, so the D-instanton gives rise to hard
scattering amplitudes of order $e^{-O(1/g_{\rm st})}$.  In fact the
effective action~\ee{seff} is essentially a local exponential in
spacetime, which suggests that the theory may have severe problems in
the ultraviolet.  This may not be fatal, as will argue in the
next section that the divergences in the D-instanton amplitude cancel.
In any case, we again note that we are more interested in the
combinatoric properties of the boundaries than in the particular case of
Dirichlet boundary conditions.

\section{Divergences and Anomalies}

Dirichlet boundaries lead to a string divergence of a rather severe
sort, arising when a vertex operator or group of vertex
operators approaches the boundary.  An example is shown in figure~1a,
and an equivalent representation in terms of a degenerating strip in
figure~1b.
\begin{figure}
\epsfbox{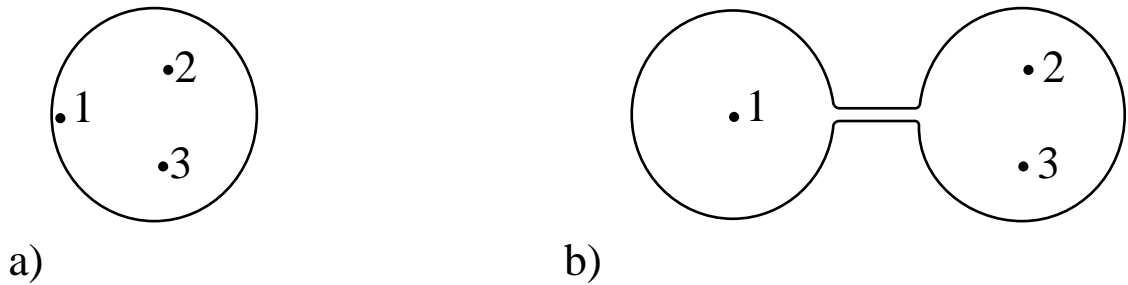}
\caption{a) Dirichlet disk amplitude with three closed-string vertex
operators, with $V_1$ approaching the boundary.  b) Conformally
equivalent picture, two disks joined by a degenerating strip.}
\end{figure}
The divergence comes from the $L_0 = 1$ state $\alpha^\mu
|0\rangle$ in the Hilbert space of the open string with
endpoints fixed at identical positions, and is of the form
\eq
<V_1 B^\mu>_{D_2,X} <V_2 V_3 B_\mu>_{D_2,X} \int_0 \frac{dt}{t} \e{div1}
\eqe
where
\eq
B^\mu = \oint_B ds\, \del_n X^\mu \e{cvo}
\eqe
is the vertex operator for this state and $t \to 0$ is the limit of
moduli space.  There is also a tachyon divergence $\int_0 dt/t^3$, which
can as usual be defined by analytic continuation.  More generally, this
divergence appears whenever a strip with coincident endpoints
degenerates; this includes the
short-distance divergence noted above.

To deal with this divergence we recall the
Fischler-Susskind principle\cite{FS}, that physically sensible
quantities are free of divergences.  Note that the vertex
operator~\ee{cvo}
is just the collective coordinate~\ee{vo} for the position of the
D-instanton.
In the original Fischler-Susskind
situation~\cite{FS}, to cancel divergences it was necessary to expand around
the correct background configuration.  The present case is different,
because there is no momentum-dependence in the divergence, and because
there is no `correct' value of the collective coordinate---rather, we
must integrate over it.  Note now that the divergence~\ee{div1} can be
put in the form
\eq
\frac{\del}{\del X_\mu} <V_1>_{D_2,X}\frac{\del}{\del
X^\mu} <V_2 V_3>_{D_2,X} \int_0 \frac{dt}{t}
\e{div2}
\eqe
This is not a total derivative as it stands, but there are two other
divergent amplitudes, shown in figures~2a and~2b,
\begin{figure}
\epsfbox{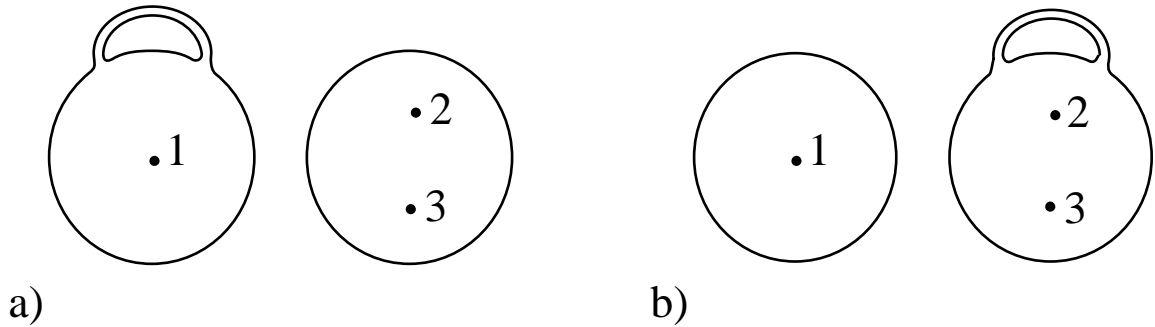}
\caption{a,b) Amplitudes which cancel the divergence and anomaly of
figure~1.  In each amplitude, there are three boundaries at a common
position $X^\mu$.}
\end{figure}
which combine with
the divergence~\ee{div3} to give
\eq
\frac{1}{2}\frac{\del}{\del X_\mu} \frac{\del}{\del
X^\mu} <V_1>_{D_2,X} <V_2 V_3>_{D_2,X} \int_0 \frac{dt}{t}
\e{div3}
\eqe
which now vanishes upon integration (the factor of $\frac{1}{2}$
correctly accounts for the symmetry of the annulus).  This generalizes
directly to all other divergences involving this intermediate state.
As in the original Fischler-Susskind mechanism, summing over
topologies has produced a finite result.  Pretty smart, them
strings.  Again, this mechanism depends in a essential way on the
inclusion of multiple boundaries at the same D-instanton; the
cancellation is between $n_a = 1$ in figure~1 and $n_a = 3$ in
figure~2.\footnote
{In the $n_a \equiv 1$ Dirichlet theory there are therefore uncancelled
divergences, about which Green has put forward a very interesting
proposal\cite{Ginf}.  He interprets the state~\ee{cvo}
as a Lagrange multiplier, whose most notable
effect is to remove the dilaton from the string spectrum.  This is
completely different from the role of this state in our approach,
but we can point to no obvious inconsistency.}

The divergence is accompanied by a conformal anomaly in the individual
graphs, because any cutoff on the integration over $t$ will rescale under a
conformal transformation.  Again this cancels between the various
topologies.  This has a curious consequence.  Comparison with
matrix model results requires a linear dilaton background.  It would
appear that Dirichlet conditions are inconsistent with a
gradient of the dilaton $\Phi$, because the world-sheet field
$X^\mu(\sigma)$ is not Weyl-invariant,
\eq
\delta X^\mu(\sigma) = \delta\phi(\sigma) \del^\mu \Phi(X(\sigma)).
\eqe
However, one is free to take the vertex operator $V_1$ to be a dilaton
with nonzero momentum, and the tree-level conformal anomaly from figure~1
is cancelled by the amplitudes of figure~2.  So the Dirichlet boundary
condition is evidently consistent with a linear dilaton background
even though it does {\it not} give rise to a conformal field theory,
by cancellation in the string loop expansion.

\section{Discussion}

We first summarize our results for the specific case of Dirichlet
boundaries, before turning to broader and more speculative issues.
We have argued on several grounds---duality, the analogy with the
D-brane, and the Fischler-Susskind mechanism, that the nature of the
theory with Dirichlet boundaries is rather different from what has
been considered previously.  A consequence is that the D-instanton
carries a weight $e^{-O(1/g_{\rm st})}$, so that the Dirichlet theory
is identical to the usual closed string theory to all orders of
perturbation theory.  The difference is of the same order as the
indeterminacy in the perturbation series itself\cite{Shenker}.

Are we to interpret this as a nonperturbative ambiguity in the
theory, that the usual closed string and the Dirichlet string theory
are two different theories?
Our speculation is that there is not a large nonperturbative
ambiguity in string theory.\footnote
{For example, in the $d=1$ matrix model, we have found that most of
the nonperturbative ambiguity is removed by considerations of
causality\cite{jpprep}.}  Rather,
given an exact nonperturbative formulation of string theory, the
leading nonperturbative effects would be given as a sum over specific
boundary types.  Note that the Dirichlet boundary condition is only
one possible example.  The condition for a boundary state $|B\rangle$
to define a conformal field theory is
\eq
(L_n - \tilde L_{-n}) |B\rangle
= 0. \e{repar}
\eqe
There are many such states, one for each primary field in the
theory\cite{CMNP2,IO}.  There are other consistency conditions as
well\cite{Lew}---for example, there must be a Hilbert space
interpretation in the open string channel, in order that amplitudes
properly factorize---and we do not know the general solution.

In $d = 2$ string theory\cite{d1refs}, the stringy nonperturbative
effects are understood in terms of tunneling of matrix model
fermions.\footnote{For a recent discussion see ref.~\cite{tunn}.}
This would
correspond to effective interactions of the general form
\eq
e^{-O(1/g_{\rm st})}e^{i \{ S_{R,L} (q) - S_{R,L} (q')
\}/\sqrt{2\pi}},
\eqe
where $S(q)$ is the canonically normalized tachyon field.  This is of
the general form~\ee{seff}, the exponent in the second term being of
order~$g_{\rm st}^0$.  It is therefore consistent with our more
general speculation about boundaries.  It does not appear to
correspond specifically to a Dirichlet boundary condition,
however, as we see no
way to obtain the necessary $i$ in the exponent from these.
Also, the Dirichlet amplitudes have a very simple momentum dependence,
while a more complicated structure, including leg poles, seems to
arise in the matrix model.\footnote{For a recent discussion of leg
poles, with references to earlier work, see ref.~\cite{NP}.}

Dirichlet boundaries for the type II superstring were discussed in
ref.~\cite{gII}, but boundaries in the heterotic string are problematic.
The condition~\ee{repar} has no natural extension to a chiral
algebra; perhaps it must simply be replaced by BRST invariance
(together with the other conditions mentioned above).

To conclude let us emphasize a possible general lesson.  This is the
nonperturbative breakdown of the world-sheet---through the appearance
of holes (which has been also discussed in a different
context\cite{AW}) and in the extrinsic nature of the D-instanton.

\section*{Acknowledgements}

I would like to thank S. Chaudhuri, M. Green, M.
Natsuume, and E. Smith for discussions.  This work was supported in
part by National Science Foundation grants PHY-89-04035 and
PHY-91-16964.


\newpage

\begin{thebibliography}{100}
\bibitem{Shenker} S. Shenker, in {\it Random Surfaces and Quantum Gravity},
eds. O. Alvarez, E. Marinari, and P. Windey (Plenum, 1991).
\bibitem{parton1} M. B. Green, Phys. Lett. {\bf B69}, 89 (1977); {\bf B201},
42 (1988); {\bf B282}, 380 (1992).
\bibitem{parton2} Z. Yang, ``Asymptotic Freedom and Dirichlet String Theory,''
preprint UR-1288, ER-40685-737, hep-th/9211092 (1992)\semi
M. Li, ``Dirichlet Strings,'' preprint Brown-HET-915, hep-th/9307122 (1993).
\bibitem{parton3} M. B. Green and J. Polchinski, ``Summing over World-Sheet
Boundaries,'' preprint DAMTP/94-38, NSF-ITP-94-52, hep-th/9406012 (1994).
\bibitem{CMNP} J. H. Schwarz, Nucl. Phys. {\bf B65}, 131 (1973)\semi
E. F. Corrigan and D. B. Fairlie, Nucl. Phys. {\bf B91}, 527 (1975)\semi
M. B. Green, Nucl. Phys. {\bf B103}, 333 (1976)\semi
A. Cohen, G. Moore, P. Nelson, and J. Polchinski,
Nucl. Phys. {\bf B267}, 143 (1986); {\bf B281}, 127 (1987).
\bibitem{CMNP2} A. Cohen, G. Moore, P. Nelson, and J. Polchinski,
in {\it Unified String Theories}, eds. M. Green and
D. Gross (World Scientific, Singapore, 1986).
\bibitem{Siegel} W. Siegel, Nucl. Phys. {\bf B109}, 244 (1976).
\bibitem{DLP} J. Dai, R. Leigh, and J. Polchinski, Mod. Phys. Lett. {\bf A4},
2073 (1989)\semi
R. Leigh, Mod. Phys. Lett. {\bf A4}, 2767 (1989).
\bibitem{Gdual} M. B. Green, Phys. Lett. {\bf B266}, 325 (1991).
\bibitem{tad} B. Grinstein and M. Wise, Phys. Rev. {\bf D35}, 655 (1987);
3285(E)\semi
M. R. Douglas and B. Grinstein, Phys. Lett. {\bf B183}, 52 (1987); {\bf B187},
442 (1987)(E)\semi
J. Liu and J. Polchinski, Phys. Lett. {\bf B 203}, 39 (1988).
\bibitem{BD} T. Banks and M. Dine, ``Coping with Strongly Coupled String
Theory,'' preprint RU-50-94, SCIPP 94/12, hep-th/9406132.
\bibitem{FS} W. Fischler and L. Susskind, Phys. Lett. {\bf B171}, 383 (1986);
{\bf B173}, 262 (1986).
\bibitem{Ginf} M. B. Green, Phys. Lett. {\bf B302}, 29 (1993).
\bibitem{jpprep} J. Polchinski, in preparation.
\bibitem{IO} N. Ishibashi, Mod. Phys. Lett. {\bf A4}, 251 (1989)\semi
J. Cardy, Nucl. Phys. {\bf B324}, 581 (1989).
\bibitem{Lew} D. Lewellen, Nucl. Phys. {\bf B372}, 654 (1992).
\bibitem{d1refs} D. J. Gross and N. Miljkovi\'c, Phys. Lett. {\bf B238},
217 (1990)\semi
E. Br\'ezin, V. A. Kazakov, and A. B. Zamolodchikov, Nucl. Phys.
{\bf B333}, 673 (1990)\semi
P. Ginsparg and J. Zinn-Justin, Phys. Lett. {\bf B240}, 333 (1990)\semi
G. Parisi, Phys. Lett. {\bf B238}, 209 (1990).
\bibitem{tunn} R. Brustein, M. Faus, and B. A. Ovrut, ``Instanton Effects in
Matrix Models and String Effective Lagrangians,'' preprint CERN-TH. 7301/94,
UPR-608T, hep-th/9406179 (1994).
\bibitem{NP} M. Natsuume and J. Polchinski, ``Gravitational Scattering in the
$c=1$ Matrix Model,'' preprint NSF-ITP-94-19, hep-th/9402156 (1994),
to appear in Nucl. Phys. B.
\bibitem{gII} M. B. Green, ``Pointlike States for Type 2b Superstrings,''
preprint DAMPT 94/19, hep-th/9403040 (1994).
\bibitem{AW}  J. Atick and E. Witten, Nucl. Phys. {\bf B310}, 291 (1981).
\end{thebibliography}
\end{document}